\documentclass[prl,reprint,superscriptaddress]{revtex4-2}
\usepackage{amsmath}		
\usepackage{amssymb}		
\usepackage{mathtools}
\usepackage{mathrsfs}
\usepackage{bm}
\usepackage{graphicx}		
\usepackage{subfigure}		
\usepackage{physics}		
\usepackage{tensor}			
\usepackage{verbatim}
\usepackage{hyperref}
\usepackage{xcolor}
\usepackage[normalem]{ulem}
\usepackage{braket}

\usepackage[textcolor=red,textsize=small,textwidth=0.7in,backgroundcolor=white,bordercolor=white]{todonotes}

\newcommand{\vz}{{\bm z}}
\newcommand{\veta}{{\bm\eta}}
\newcommand{\ee}{\mathrm{e}}
\newcommand{\ez}{{\bm{e}_z}}
\newcommand{\vq}{{\bm q}}

\newcommand{\ii}{{\mathrm i}}

\newcommand{\vx}{{\bm x}}
\newcommand{\vA}{{\bm A}}

\newcommand{\vecb}{{\bm b}}

\newcommand{\vk}{{\bm k}}

\newcommand{\vbeta}{{\bm\beta}}

\begin{document}
\title{Generalizing the composite fermion theory for fractional Chern insulators}

\author{Hao Jin}

\affiliation{International Center for Quantum Materials, Peking
	University, Beijing 100871, China}

\author{Junren Shi}
\email{junrenshi@pku.edu.cn}

\affiliation{International Center for Quantum Materials, Peking
	University, Beijing 100871, China}

\affiliation{Collaborative Innovation Center of Quantum Matter,
	Beijing 100871, China}

\begin{abstract}
  We propose a generalized composite fermion (CF) theory for fractional Chern insulators (FCIs) by adapting the quantum mechanics approach of CFs.
The theoretical framework naturally produces an effective CF Hamiltonian and a wavefunction ansatz, and the Bloch band characteristics of FCIs determine effective scalar and vector potentials experienced by CFs.
Our analysis clarifies the construction of CF wavefunctions and state counting in CF phase space, which is subject to a density-of-states correction for filling factors $|\nu| \neq 1/2$.
We apply the theory to study the $\nu=-2/3$ FCI state of the twisted bilayer MoTe$_2$ system, modeling it as either a $1/3$-filled electron band or a $2/3$-filled hole band.
While both CF models exhibit trends and features consistent with exact diagonalization results, the electron-based model shows better agreement.
Furthermore, we find that the FCI phase transition coincides with a topological phase transition in unoccupied CF $\Lambda$-bands.
\end{abstract}

\maketitle

\paragraph{Introduction.---}
Fractional Chern insulators (FCIs)~\cite{neupert_fractional_2015,CRPHYS_FQH_Chern,Bergholtz_topological_flat_band,Ju_FQAH} are Bloch systems that exhibit the fractional quantum Hall (FQH) effect~\cite{Tsui_FQHE_experiment,Stormer_FQHE_review}.
It has been predicted that interacting electrons in a fractionally filled flat band, whether topologically trivial~\cite{Simon_FCI_zero,Lin_FCI_zero_Chern_number} or non-trivial~\cite{neupert_fractional_2015,CRPHYS_FQH_Chern,Bergholtz_topological_flat_band}, could form the novel correlated state of an FCI.
Recently, experimental evidences of FCI states are observed
in twisted bilayer MoTe$_2$~\cite{cai2023signatures,park2023,xu2023,zeng2023} and multilayer graphene~\cite{Lu_FCI_multilayer_grphene}. 
Unlike Landau-level (LL) systems, Bloch band characteristics, such as Chern number~\cite{Wu_model_all_chern}, Berry curvature, quantum metric tensor~\cite{Roy_band_geometry}, and dispersion~\cite{Grushin_enhancing_stability}, influence the stability and properties of an FCI.
These complexities present a challenge for FQH theories originally formulated for LL systems, which must be adapted for FCIs and critically tested.


The composite fermion (CF) theory~\cite{jain2007composite}, the \emph{de facto} standard FQH theory, has been the focus of the generalization.
Two attempts have been made so far.
An approach based on the Chern-Simons (CS) CF picture~\cite{Lopez_FQHE_CS,HLR-theory} obtains effective CF Hamiltonians by modifying the phases of hopping integrals in FCI tight-binding models~\cite{Lu_FCI_MoTe2_CF,Sohal_CS_CF}.
On the other hand, Murthy and Shankar generalize the Hamiltonian approach of CFs~\cite{Murthy_Hamiltonian_FQHE} by expressing the projected density operators of an FCI in terms of their LL counterparts~\cite{Murthy_Hamiltonian_FCI}.

In this Letter, we present a generalization based on a quantum mechanics approach of CFs~\cite{shi2023quantum}.
Motivated by Read's dipole picture of CFs~\cite{Read_dipole_picture}, the approach establishes a CF wave function ansatz and effective CF quantum mechanics on a common ground.
It stipulates how to construct CF Hamiltonians based on specific physical contexts and map CF states deduced from the Hamiltonians to physical states.
It not only reproduces established results of the standard CF theory for the lowest LL but also successfully resolves anomalies encountered in the first excited LL~\cite{Jin_non-quasiconvex}, demonstrating its adaptability for varied physical contexts.
Exploring its adaptation for FCIs thus constitutes a logical next-step.

\begin{figure}[t]
	\centering
	\includegraphics[width=0.75\linewidth]{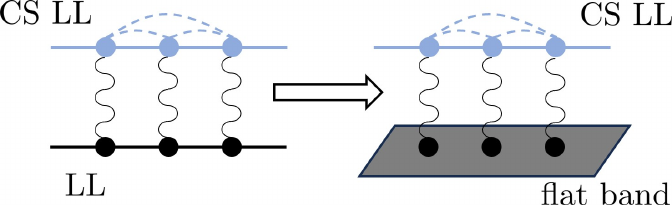}
	\caption{\label{fig:dipole_scheme}
          Adaptation of the generalized dipole model of CFs for FCIs.
          Left: The model for an LL.
          The black line and dots represent an LL and carriers occupying it, respectively.
          Vortices (blue dots) forming a half-filled bosonic Laughlin state are treated as independent particles occupying a fictitious CS LL (blue line).
          Carriers bind to vortices to form CFs, as indicated by wavy lines.
          Right: The model is adapted for FCIs by replacing the LL with a flat band.}
\end{figure}

\paragraph{Adapting the quantum mechanics approach.---}
The quantum mechanics approach of CFs is based on the generalized dipole picture of CFs illustrated in Fig.~\ref{fig:dipole_scheme}~\cite{shi2023quantum}. 
In this picture, carriers (electrons or holes) confined in a physical LL bind to vortices forming a bosonic Laughlin state at half-filling.
Under the mean field approximation, vortices are treated as independent particles residing in a fictitious LL induced by an emergent CS field. 
The picture can be adapted for FCIs straightforwardly by substituting the LL with a flat band, as demonstrated on the right of Fig.~\ref{fig:dipole_scheme}. 

The picture leads to a wave function ansatz that obtains a physical wave function $\Psi_\text{e}$ by projecting a CF state $\ket{\Psi_{\mathrm{CF}}}$ onto the half-filled bosonic Laughlin state of vortices $\ket{\Psi_\text{v}^\text{L}}$~\cite{read1998,shi2023quantum,hu2024}: $\Ket{\Psi_{\text{e}}}=\Braket{\Psi_{\text{v}}^{\text{L}}|\Psi_{\text{CF}}}_{\mathrm{v}}$, where the overlap is taken partially over the fictitious vortex degrees of freedom.
Ref.~\onlinecite{shi2023quantum} establishes that the ansatz is equivalent to Jain's wave function ansatz, which underlies the great success of the CF theory in LL systems.
It is straightforward to apply the ansatz to FCIs as well.

The picture also implies the modeling of CF energy, which is supported by a microscopic analysis~\cite{shi2023quantum}.
For an ideal LL system, it only consists of the binding energy arising from the Coulomb attraction between a charge carrier and the charge void induced by the vortex associated with the carrier~\cite{Read_dipole_picture,shi2023quantum}.
Unlike an LL, a flat band generally has a dispersion $\epsilon_{\bm k_{c}}$ that depends on the carrier quasi-wavevector $\bm{k}_{c}$.
Thus, for an FCI, the energy of a CF comprises two parts, given by $\epsilon_{\text{CF}}=\epsilon_{\bm{k}_{c}}+\hbar^{2}\abs{\vz-\veta}^{2}/2m^{*}l_{c}^{2}l_{v}^{2}$,
where the second term is a harmonic binding potential with $\bm z$ and $\bm \eta$ denoting the coordinates of the carrier and the vortex in a CF, respectively.
For the lowest LL, it has been shown that the binding energy can be well approximated as a harmonic potential~\cite{shi2023quantum,Jin_non-quasiconvex}.
Here, we assume that this still holds true for FCIs, and parameterize the coefficient analogously, with $m^*$ being the CF effective mass, and $l_{v}=1/\sqrt{4\pi\rho_{c}}$ representing the CS magnetic length, which is determined by the carrier density $\rho_{c}$.
In analogy to the magnetic length in an LL, we define a carrier length scale $l_{c}\equiv \sqrt{A_{p}/2\pi}$, where $A_{p}$ is the primitive cell area of the system.

\paragraph{CF effective Hamiltonian.---}
The effective CF Hamiltonian is derived by evaluating the matrix elements of $\epsilon_{\mathrm{CF}}$ in a chosen basis set~\cite{shi2023quantum}.
For an FCI, it is convenient to use the Bloch basis of the flat band for a carrier and the magnetic Bloch basis of the CS LL for a vortex.
The CF wave function, $\varphi(\bm k_{c},\bm k_{v})$, thus depends on the quasi-wavevectors $\vk_{c}$ and $\vk_{v}$ of the electron and vortex, respectively, and is determined by the effective CF Hamiltonian given by~\cite{shi2023quantum,supp}
\begin{align}
\hat{H} & =\frac{1}{2m^{*}}\hat{\bm{\Pi}}^{2}+U_{\bm{k}_{c}},\label{eq:HCF}\\
\hat{\bm{\Pi}} & \equiv\frac{\hbar}{l_{c}l_{v}}\qty[\qty(\ii\nabla_{\vk_{c}}+\vA_{\vk_{c}})-\qty(\ii\nabla_{\vk_{v}}+\bm{A}_{\bm{k}_{v}}^{(\mathrm{v})})],\\
U_{\bm{k}_{c}} & \equiv\epsilon_{\vk_{c}}+\frac{\hbar^{2}}{2m^{*}l_{c}^{2}l_{v}^{2}}\Tr\mathbb{G}_{\vk_{c}},
\end{align}
where $\vA_{\vk_{c}}$ and $\mathbb{G}_{\vk_{c}}$ are the Berry connection and quantum metric tensor of the flat band~\cite{Roy_band_geometry}, respectively, and $\bm A_{\bm k_{v}}^{(\mathrm{v})}$ is the Berry connection of the CS band. 
The Berry connection of a flat band with Chern number $C=-1$ can be decomposed as $\bm A_{\bm k_{c}}=-\frac{l_{c}^{2}}{2}\bm e_{z}\times \bm k_{c} + \delta\bm A_{\bm k_{c}}$,
with $\delta\bm A_{\bm k_{c}}$ being periodic in $\bm k_{c}$, where $\bm{e}_{z}$ denotes the unit vector normal to the system plane.
On the other hand, the CS LL is interpreted as an ideal flat band~\cite{Wang_ideal_Chern_band} with the Berry connection $\vA_{\vk_v}^{(\mathrm{v})}=\frac{l_{v}^{2}}{2}\ez\times\vk_{v}$, corresponding to a uniform Berry curvature and Chern number $C_{v} = 1$.

A CF wavefunction in the wavevector representation must be quasi-periodic with respect to the basis vectors of both the carrier and vortex reciprocal primitive cells (RPCs).
The requirement imposes the condition:
\begin{equation}
\hat{t}_{c}(\bm{b}_{i})\varphi(\bm{k}_{c},\bm{k}_{v})=\hat{t}_{v}(\bm{\beta}_{i})\varphi(\bm{k}_{c},\bm{k}_{v})=\varphi(\bm{k}_{c},\bm{k}_{v}),
\label{eq:phi_pbc_k}
\end{equation}
where $\hat{t}_{c}(\bm{b}_{i})\equiv\exp(\bm{b}_{i}\cdot\nabla_{\bm{k}_{c}}-\mathrm{i}l_{c}^{2}\bm{k}_{c}\land\bm{b}_{i}/2)$ and $\hat{t}_{v}(\bm{\beta}_{i})\equiv\exp(\bm{\beta}_{i}\cdot\nabla_{\vk_v}+\mathrm{i}l_{v}^{2}\bm{k}_{v}\land\bm{\beta}_{i}/2)$ are reciprocal translation operators of carriers and vortices, respectively, with $\bm b_{i}$ and $\vbeta_i$ (for $i=1,2$) being the basis vectors of respective RPCs.
We choose $\bm\beta_1=(l_c^2/l_v^2)\vecb_1$ and $\bm\beta_2=\vecb_2$. 

\paragraph{Twisted bilayer MoTe$_{2}$ system.---}
We apply our CF model to the $\nu=-2/3$ FCI state observed in the twisted bilayer MoTe$_2$ system~\cite{cai2023signatures,park2023,xu2023,zeng2023}.
To determine flat band quantities required for constructing Eq.~\eqref{eq:HCF}, we adopt the effective electron model introduced in Refs.~\onlinecite{Wu_model_Hamiltonian,Yu_model_Hamiltonian}, with parameters determined in Ref.~\onlinecite{Wang_MoTe2_ED}.
For the CF effective mass, we adopt the value commonly used for LL systems~\cite{HLR-theory,Hu_Kohn-Sham_FQHE}, given by $\hbar^{2}/m^{\ast}\approx0.3e^{2}l_{c}/4\pi\varepsilon_{0}\varepsilon_{r}$, where $\varepsilon_r$ ($\varepsilon_0$) denotes the relative (vacuum) permittivity. 

We could have two alternative CF models describing the specific FCI state, depending on whether the system is treated as a $1/3$-filled electron band or a $2/3$-filled hole band.
Since the current theoretical framework lacks an \emph{a priori} principle for selecting a preferred model, we construct and evaluate both.

The flat band dispersion $\epsilon_{\bm k_{c}}$ entering into Eq.~\eqref{eq:HCF} differs from that determined from the effective model because the model assumes a fully-filled flat band, whereas the FCI is $1/3$-filled.
To account for the difference, we apply the Hartree-Fock approximation to determine the electron dispersion at $1/3$:
\begin{equation}
\epsilon_{\bm{k}}^{(1/3)}\approx\epsilon_{\bm{k}}^{(1)}-\frac{1}{A_{\mathrm{tot}}}\sum_{\bm{k}^{\prime}}\left(V_{\vk\vk',\vk'\vk}-V_{\vk\vk',\vk\vk'}\right)n_{\vk'}^{\mathrm{h}},\label{eq:epsilonke}
\end{equation}
where $\epsilon_{\bm k}^{(1)}$ is the dispersion from the effective model, $A_{\mathrm{tot}}$ is the total area of the system, $V_{\vk\vk^{\prime},\vk_{1}^{\prime}\vk_{1}}$ denotes the matrix element of the interaction adopted in Ref.~\onlinecite{Wang_MoTe2_ED} for exact diagonalization (ED) calculations, and $n_{\vk^{\prime}}^{\mathrm{h}}$ is the hole occupation number at $\vk^{\prime}$.
For the electron- (hole-) based CF model, $\epsilon_{\bm k_{c}} \equiv \epsilon_{\bm k_{c}}^{(1/3)}$ ($-\epsilon_{\bm k_{c}}^{(1/3)}$), and $n_{\vk^{\prime}}^{\mathrm{h}}\equiv 1 - n_{c}(\bm k^{\prime})$ ($n_{c}(\bm k^{\prime})$) is determined self-consistently from the carrier density $n_{c}(\bm k^{\prime}) = (2\pi)^{2} A_{\mathrm{tot}}^{-1}\sum_{i}\int_{\mathrm{RPC_{\mathrm{v}}}} \mathrm{d}\bm k_{v} |\varphi_{i}(\bm k^{\prime}, \bm k_{v})|^{2}$, where the summation over the CF state index $i$ includes all occupied states, and the $\bm k_{v}$ integral is taken over the vortex RPC~\cite{shi2023quantum}.

\paragraph{CF Bloch bands and $\Lambda$-bands.---}
To solve the effective CF Hamiltonian Eq.~\eqref{eq:HCF}, we map it to a Hofstadter Hamiltonian defined in a fictitious $\bm x$-space by introducing a new set of variables~\footnote{
  Throughout this work, we interchangeably use the two sets of variables $(\bm k_{c},\bm k_{v})$ and $(\bm k, \bm x)$.
  Variable substitutions are implied where appropriate.}: 
\begin{align}
\vx & \equiv-l_cl_{v}\boldsymbol{e}_{z}\times\vk_{c},\label{eq:x_def}\\
\vk & \equiv\vk_{c}+\vk_{v},\label{eq:k_def}
\end{align}
and defining
$\hat{H}'\equiv\exp[\ii(l_{v}/2l_c)\vk\cdot\vx]\hat{H}\exp[-\ii(l_{v}/2l_c)\vk\cdot\vx]$.
We have:
\begin{align}
\hat{H}^{\prime} & =\frac{\left[-\mathrm{i}\hbar\bm{\nabla}_{\bm{x}}+e\bm{\mathcal{A}}(\bm{x})\right]^{2}}{2m^{\ast}}+U(\bm{x}),\label{eq:H'}\\
e\boldsymbol{\mathcal{A}}\qty(\vx) & \equiv-\hbar\boldsymbol{e}_{z}\times\qty(\frac{\sigma}{2l^{2}}\vx-\frac{\delta{\vA}_{\vk_{c}}}{l_cl_{v}}),\label{eq:vecA_def}
\end{align}
where $l$ denotes the effective magnetic length,
defined as $\sigma l^{-2}\equiv l_{c}^{-2}-l_v^{-2}$ with $\sigma=1$ ($-1$) for $\nu_{\rm c} < 1/2$ ($>1/2$), $U(\bm x)\equiv U_{\bm k_{c}}$ and $\mathcal{A}(\bm x)$ are effective scalar and vector potentials, respectively.
In the resulting Hamiltonian, the dispersion, Berry connection, and quantum metric tensor of the flat band collectively determine the scalar and vector potentials experienced by CFs.
It possesses two competing periodicities: one derived from the underlying periodicity of the flat band, and the other arising from a uniform effective magnetic field characterized by $l$.  



\begin{figure}[b]
	\centering
	\includegraphics[width=\linewidth]{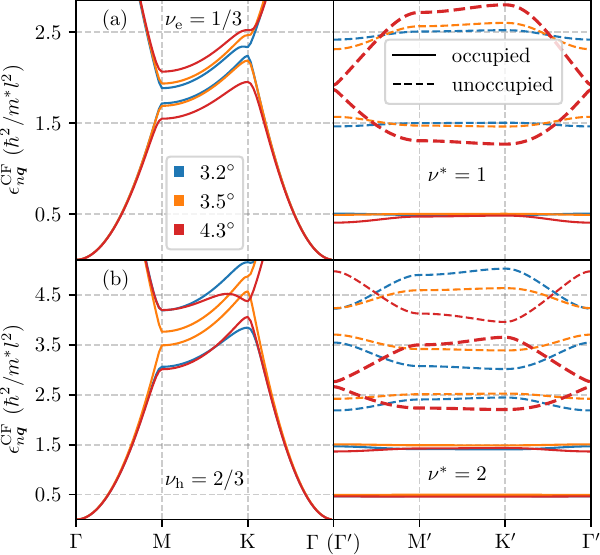}
	\caption{
		Dispersions of CF Bloch bands prior to the magnetic  splitting (left) and $\Lambda$-bands (right) of a twisted bilayer MoTe$_{2}$ with a $1/3$-filled mori\'{e} valence band at various twist angles $\theta$ and $\epsilon_r=15$. 
		The system is treated either as (a) a $1/3$-filled electron band
		and (b) a $2/3$-filled hole band. 
                Solid (dashed) lines in the right panels indicate (un-)occupied bands. 
		The positions of the high symmetry points are defined in Fig.~\ref{fig:BZ_MBZ}.
              }
	\label{fig:CF_dispersion}
\end{figure}




\begin{figure}[b]
	\centering
	\includegraphics[width=\linewidth]{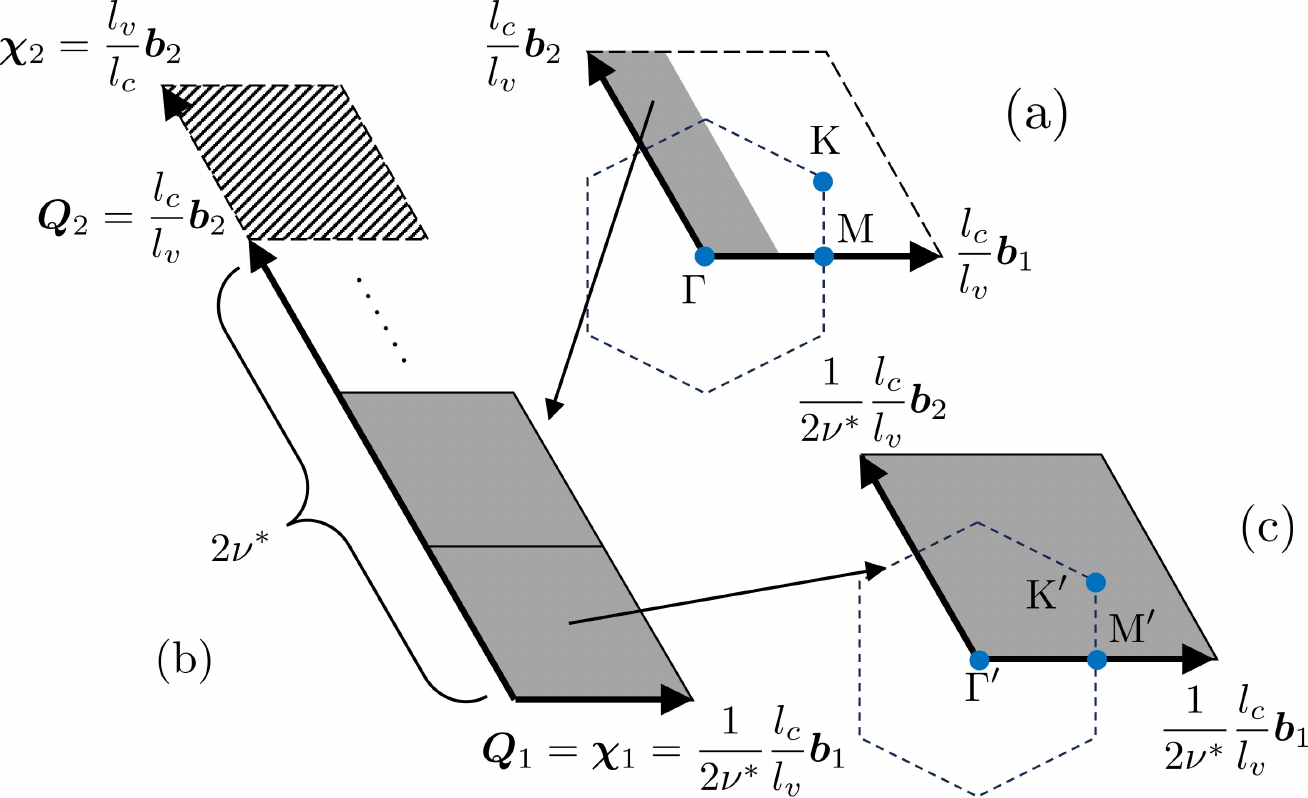}
        \caption{Definitions and relationships of various RPCs for the filling fraction $\nu_{\rm c}=\nu^{*}/(2\nu^{*}+1)$.
          (a) The RPC of the Hofstadter system prior to magnetic splitting.
          (b) The magnetic RPC of the Hofstadter system, partitioned into $2\nu^{\ast}$ subcells and shown as filled blocks.
          The RPC of the corresponding CF system is one subcell larger, as indicated by the shaded block. 
          (c) The reduced magnetic RPC corresponding to a subcell.
          For the fraction $\nu_{\rm c}=\nu^{*}/(2\nu^{*}-1)$, the relations are similar, but the CF-RPC is one-subcell smaller.
        }
	\label{fig:BZ_MBZ}
\end{figure}

By removing the uniform effective magnetic field component in Eq.~\eqref{eq:vecA_def}, we obtain a periodic system, which yields CF Bloch bands shown in the left panels of Fig.~\ref{fig:CF_dispersion}.
These bands are separated by gaps induced by the periodic effective potentials.
In the presence of the uniform effective magnetic field, the CF Bloch bands split into magnetic subbands, as shown in the right panels of Fig~\ref{fig:CF_dispersion}.
They are the FCI analogs of the $\Lambda$-levels of CFs in an LL~\cite{jain2007composite}, and are thus referred to as $\Lambda$-bands.

For the CF filling fraction $\nu_{\rm c}=\nu^*/\qty(2\nu^*\pm 1)$, $\nu^*\in\mathbb N^+$, each CF Bloch band splits into $2\nu^{*}$ $\Lambda$-bands, as the primitive cell areas corresponding to the two competing periodicities are $2\pi l_v^2$ and $2\pi l^{2}$, respectively, and $l^{2} = 2\nu^{*} l_{v}^{2}$.
To solve $H^{\prime}$, we enlarge the $\bm x$-space primitive cell by extending it $2\nu^{*}$-fold along one of its primitive vectors, $\bm\zeta_1=l_cl_v\vecb_2\times \ez$. 
The resulting magnetic RPC is reduced to $1/2\nu^*$ of the one prior to the splitting.
Moreover, each magnetic RPC can be further partitioned into $2\nu^{\ast}$ subcells, across which the $\Lambda$-bands are degenerate.
This degeneracy arises because the set of $2\nu^{\ast}$ magnetic translation operators, $\{\hat{t}(m\bm\zeta_{1}),\,m=0,\cdots,2\nu^{\ast}-1\}$ with $\hat{t}\qty(\boldsymbol{d})\equiv\exp(\bm{d}\cdot\nabla_{\vx}\pm\ii\vx\wedge\boldsymbol{d}/2l^{2})$, commutes with $\hat H^{\prime}$ and generates distinct eigenstates.
The subcell defines the reduced magnetic RPC.
The definitions of these RPCs, as well as their relationships, are shown in Fig.~\ref{fig:BZ_MBZ}.





\paragraph{CF states and phase space.---}
While each Bloch eigenfunction $\phi_{n\vq}\qty(\vx)$ for $\hat H^{\prime}$, where $n$ and $\bm q$ denote the $\Lambda$-band index and quasi-wavevector, respectively, yields an eigenfunction $\phi_{n\vq}\qty(\vx)\exp[-\ii(l_{v}/l_{c})\vk\cdot\vx/2]$ for $\hat H$, it is not a valid CF wavefunction as it does not satisfy the quasi-periodic boundary condition Eq.~\eqref{eq:phi_pbc_k}. 
We thus construct the CF wavefunction as a superposition of the $2\nu^{*}$ degenerate wavefunctions generated by the magnetic translation operators:
$\varphi_{n\vq}\qty(\vk_{c},\vk_{v})=e^{-\ii(l_{v}/2l_{c})\vk\cdot\vx}\sum_{m=0}^{2\nu^{*}-1}a_{m}\qty(\vk)\hat{t}\qty(m\bm{\zeta}_{1})\phi_{n\vq}\qty(\vx)$.
By imposing the boundary condition Eq.~\eqref{eq:phi_pbc_k}, $a_{m}(\bm k)$ can be determined up to an unimportant $\bm q$-dependent phase factor~\cite{supp}. 
The resulting CF wave function is given by:
\begin{multline}
\varphi_{n\bm{q}}\qty(\bm{k}_{c},\bm{k}_{v})\propto\ee^{-\frac{\ii}{2}\frac{l_{v}}{l_{c}}\vk\cdot\vx}\sum_{\bm{\chi}}\delta\left(\frac{l_{v}}{l_{c}}\vk-\vq-\bm{\chi}\right)\\
\times\sum_{m=0}^{2\nu^{*}-1}\ee^{\ii\pi\frac{l_{v}}{l_{c}}\left[2q_{2}-\frac{l_{v}}{l_{c}}\left(k_{2}+2m\right)\right]k_{1}}\hat{t}\qty(m\boldsymbol{\zeta}_{1})\phi_{n\vq}\qty(\vx),\label{eq:phi_FCI}
\end{multline}
%
%
where the $\bm \chi$-summation is over the reciprocal lattice generated by the CF reciprocal basis vectors $(\bm\chi_{1},\bm\chi_{2})$ shown in Fig.~\ref{fig:BZ_MBZ}(b), $\vk\equiv k_1\vecb_1+k_2\vecb_2$, and $\vq\equiv q_1\vecb_1+q_2\vecb_2$. 


Notably, the mapping from Hofstadter states to CF states is not injective.
Depending on whether $\nu_{\rm c} < 1/2$ or $\nu_{\rm c} > 1/2$, the CF-RPC, which indexes independent CF states, is either larger or smaller than the Hofstadter RPC, as shown in Fig.~\ref{fig:BZ_MBZ}(b).

Moreover, state counting in the CF phase space differs from the usual one that each quantum state occupies an area $(2\pi)^{2}/A_{\mathrm{tot}}$ in wavevector space~\cite{Landau_statistical_physics}.
This can be inferred from Eq.~\eqref{eq:phi_FCI}, where $\bm{q}$ is related to $\bm{k}$ by
$\vq=(l_{v}/l_{c})\vk\,\pmod{\bm{\chi}_{1},\bm{\chi}_{2}}$.
Since $\bm k$ is defined in a physical space and quantized as usual, the relation implies that each CF state occupies an area $(l_{v}/l_{c})^{2}(2\pi)^{2}/A_{\mathrm{tot}}$ in the $\bm q$-space.
The extra factor can be identified as a phase space density-of-states correction~\cite{Xiao_DOS_modification}, which is expected for a topological CF momentum manifold in the presence of an effective magnetic field~\cite{shi2017,Ji_semiclassical_CF}.
Taking this into account, CFs with a density $\rho_{c} \equiv 1/4\pi l_{v}^{2}$ occupy $\nu^{\ast}$ $\Lambda$-bands defined on the CF-RPC with a $\bm q$-space area $\bm\chi_{1}\land\bm\chi_{2}=\pi/\nu^{*}l_{c}^{2} $, as indicated in Fig.~\ref{fig:CF_dispersion}.

\begin{figure}[t]
	\centering
	\includegraphics[width=\linewidth]{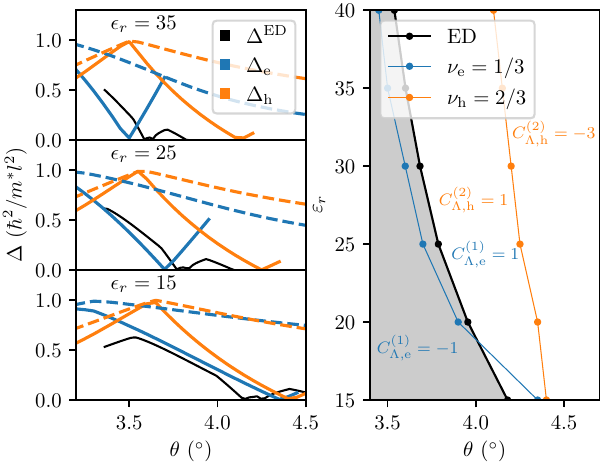}
	\caption{Energy gaps (left) and phase boundaries (right) from CF models and ED~\cite{Wang_MoTe2_ED}.
          Blue (orange) and black curves correspond to the electron- (hole-) based CF model and ED results, respectively.
          For CF models, both the mean-field gap (dashed lines) and the topological gap (solid lines) are shown. 
	The Chern numbers  $C_{\Lambda,\rm{e}}^{(1)}$ ($C_{\Lambda,\rm{h}}^{(2)}$) of the lowest unoccupied $\Lambda$-band are indicated in regions separated by the phase boundary from the electron- (hole-) based CF model. 
	}
	\label{fig:excitation_gap}
\end{figure}

\paragraph{$\Lambda$-band gaps and FCI phase transition.---}
The $\theta$-dependent mean-field CF gap, which is the gap between the highest occupied and lowest unoccupied $\Lambda$-bands, is presented as dashed lines in the left panels of Fig.~\ref{fig:excitation_gap} for various $\varepsilon_r$ values and both CF models.
The ED many-body excitation gap from Ref.~\onlinecite{Wang_MoTe2_ED} is also shown.
As the CF-exciton~\cite{Kamilla_exciton_CF} constitutes the principal excitation in a CF system, the mean-field CF gap corresponds to the CF-exciton energy at a large wavevector, while the ED gap corresponds to its minimum energy~\cite{jain2007composite}.
Although not identical, their variations are expected to correlate.

The mean-field CF gap does exhibit features and trends consistent with ED results.
For $\varepsilon_{r}=15$, both the CF and ED gaps show non-monotonic $\theta$-dependence, with maxima close to $\theta=3.5^{\circ}$.
With increasing $\varepsilon_{r}$, the $\theta$-dependence of the ED gap shifts leftward.
We find that the CF gap traces the trend, with the electron-based CF model aligning more closely with the ED results.
On the other hand, although the CF gap is mildly suppressed at large twist angles, it remains open when the ED gap closes and the FCI phase becomes destabilized.

Notably, the FCI phase transition seems to coincide with a topological transition of the lowest unoccupied $\Lambda$-band. 
In the left panels of Fig.~\ref{fig:excitation_gap}, we plot the gap between the lowest unoccupied $\Lambda$-band and the band above it for various $\varepsilon_{r}$ values, shown as solid lines and referred to as the CF topological gap.
As $\theta$ varies, the gap closes and reopens, accompanied by a change in the Chern number of the lowest unoccupied $\Lambda$-band from $-1$ ($1$) to $1$ ($-3$) for the electron- (hole-) based CF model, indicating the occurrence of a topological transition of the band.
The phase boundary of the topological transition inferred from the electron-based model closely aligns with the FCI phase boundary, as evident in the right panel of Fig.~\ref{fig:excitation_gap}.

The coincidence suggests that the FCI phase transition may be driven by changes in CF-excitons, as the properties of the bands involved in exciton formation are known to influence their nature and binding energy~\cite{kallin1984,Zhou_Berry_exciton}.
This is left for future investigation.

\paragraph{Discussion.---}
In summary, we generalize the CF theory and apply it to the MoTe$_{2}$ system, successfully reproducing trends and features observed in ED calculations.
However, the approach should be regarded as preliminary, as it does not consider the periodic potentials experienced by vortices, which are expected based on a microscopic consideration similar to that in Ref.~\onlinecite{shi2023quantum}.
This limitation may be responsible for the large discrepancy between the electron- and hole-based CF models.

\paragraph{Acknowledgments.---}
We thank Chong Wang for sharing the raw data from Ref.~\onlinecite{Wang_MoTe2_ED} and useful discussions.
This work is supported by the National Key R\&D Program of China under Grant No. 2021YFA1401900 and the National Science Foundation of China under Grant No. 12174005.

Near the completion of this work, we became aware of a related study by Hu et al.~\cite{hu2025}, which explores similar topic.
Our work complements their findings by focusing on the construction and topological transitions of $\Lambda$-bands and their implications for FCI stability.

\paragraph{Data availability.---}
The data that support the findings of this Letter are openly available \cite{open_data}, 
except for the ED data from Ref.~\onlinecite{Wang_MoTe2_ED}.

\bibliography{ref}
\end{document}


\title{Supplemental Material}
\author{Hao Jin}

\affiliation{International Center for Quantum Materials, Peking University,
	Beijing 100871, China}

\author{Junren Shi}

\affiliation{International Center for Quantum Materials, Peking University,
	Beijing 100871, China}

\affiliation{Collaborative Innovation Center of Quantum Matter, Beijing 100871,
	China}

\maketitle
\onecolumngrid

\section{CF effective Hamiltonian}

In order to derive the CF effective Hamiltonian $\hat H$, we evaluate the matrix elements of the CF energy $\epsilon_{\text{CF}}=\epsilon_{\bm{k}_{c}}+\hbar^{2}\abs{\vz-\veta}^{2}/2m^{*}l_{c}^{2}l_{v}^{2}$ in the basis set $\qty{\ket{\vk_c,\vk_v}=\ket{\vk_c}_{\rm c}\otimes\ket{\vk_v}_{\rm v}}$, where $\ket{\vk_c}_{\rm c}$ is the Bloch basis of the flat band and $\ket{\vk_v}_{\rm v}$ is the magnetic Bloch wavefunction of the CS-LL.

A CF state $\ket{\psi_\text{CF}}$ can in general be written as:
\begin{equation}
	\ket{\psi_\text{CF}}=\int\dd[2]{\vk_c}\dd[2]{\vk_v}\varphi\qty(\vk_c,\vk_v)\ket{\vk_c,\vk_v},
\end{equation}
the CF effective Hamiltonian $\hat H$ is defined by
\begin{equation}
\hat{H}\varphi(\vk_{c},\vk_{v})\equiv\int\mathrm{d}\bm{k}_{c}^{\prime}\mathrm{d}\bm{k_{v}^{\prime}}\mel{\vk_{c},\vk_{v}}{\epsilon_{\text{CF}}}{\vk_{c}^{\prime},\vk_{v}^{\prime}}\varphi(\vk_{c}^{\prime},\vk_{v}^{\prime}).
\end{equation}

For the electron degree of freedom, relevant matrix elements have been evaluated in Ref.~\onlinecite{shi2023quantum}:
\begin{align}
	\mel{\vk_c}{\epsilon_{\bm{k}_{c}}}{\vk_c'}_{\rm c}&=\epsilon_{\bm{k}_{c}}\delta\qty(\vk_c-\vk_c'),\\
	\mel{\vk_c}{\vz}{\vk_c'}_{\rm c}&=\qty(\ii\nabla_{\vk_c}+\vA_{\vk_c})\delta\qty(\vk_c-\vk_c'),\\
	\mel{\vk_c}{\vz^2}{\vk_c'}_{\rm c}&=\qty[\qty(\ii\nabla_{\vk_c}+\vA_{\vk_c})^2+\Tr\mathbb G_{\vk_c}]\delta\qty(\vk_c-\vk_c').
\end{align}

The matrix elements for the vortex degree of freedom can be obtained analogously by 
noting that the CS-LL corresponds to a $C=1$ flat Chern band with a uniform Berry curvature and quantum metric tensor~\cite{Wang_ideal_Chern_band}:
\begin{align}
	\mel{\vk_v}{\veta}{\vk_v'}_{\rm v}&=\qty(\ii\nabla_{\vk_v}+\vA_{\vk_v}^{(\text v)})\delta\qty(\vk_v-\vk_v'),\\
	\mel{\vk_v}{\veta^2}{\vk_v'}_{\rm v}&=\qty[\qty(\ii\nabla_{\vk_v}+\vA_{\vk_v}^{(\text v)})^2+\Tr\mathbb G_{\vk_v}^{(\text v)}]\delta\qty(\vk_v-\vk_v'),
\end{align}
with $\vA_{\vk_v}^{(\text v)}=\qty(l_v^2/2)\ez\times\vk_v$ and $\Tr\mathbb G_{\vk_v}^{(\text v)}=l_v^2$.

Using these matrix elements, one can readily derive the CF effective Hamiltonian presented in the main text, omitting unimportant constant terms.

\section{Construction of CF wavefunctions}
In this section we derive Eq.~(10) of the main text. We begin with
\begin{equation}
	\varphi_{\vq}\qty(\vk_{c},\vk_{v})=e^{-\frac{\ii}{2}\frac{l_v}{l_c}\vk\cdot\vx}
	\sum_{m=0}^{2\nu^{*}-1}a_{m}\qty(\vk)\hat{t}\qty(m\bm{\zeta}_{1})\phi_{\vq}\qty(\vx),
	\label{eq:varphi_phi}
\end{equation}
where for brevity we have omitted the $\Lambda$-band index $n$ as well as the implicit dependence of the coefficients on $n$ and $\bm q$, and $\vq$ is defined in the reduced magnetic RPC (see Fig.~3 of the main text). 
For wavevectors, we perform the expansions $\vk\equiv k_1\vecb_1+k_2\vecb_2$ and $\vq\equiv q_1\vecb_1+q_2\vecb_2$.
The coefficients are thus functions of $k_{1}$ and $k_{2}$:
\begin{equation}
a_{m}\qty(\vk) \equiv a_m\qty(k_1,k_2).
\end{equation}
We also assume the CF fraction $\nu_{\rm c}=\nu^{\ast}/(2\nu^{\ast} + \sigma)$, $\sigma=\pm 1$, for which different lengths are related by
\begin{equation}
	l_c^{-2}:l_v^{-2}:l^{-2}=\qty(2\nu^*+\sigma):2\nu^*:1.
\end{equation}

The magnetic translation operator satisfies the following relation:
\begin{equation}
\hat{t}(\bm{a})\hat{t}(\bm{b})=e^{\mathrm{i}\sigma\bm{a}\land\bm{b}/2l^{2}}\hat{t}(\bm{a}+\bm{b}).
\end{equation}
For the Hofstadter system, the primitive bases are chosen as $\tilde{\bm\zeta}_{1} = 2\nu^{\ast}\bm\zeta_{1} = 2\nu^{\ast} l_{c}l_{v}\vecb_2\times \ez$ and $\tilde{\bm\zeta}_{2} = - l_{c}l_{v}\vecb_1\times \ez$.
The following identity holds:
\begin{equation}
\hat{t}(\tilde{\bm{\zeta}}_{i})\phi_{\bm{q}}(\bm{x})=e^{\mathrm{i}\bm{q}\cdot\tilde{\bm{\zeta}}_{i}}\phi_{\bm{q}}(\bm{x}),\,\, i=1,2.
\end{equation}

By applying the condition $\hat t_c\qty(\vecb_i)\varphi_{\vq}\qty(\vk_c,\vk_v)=\varphi_{\vq}\qty(\vk_c,\vk_v)$, we obtain the relations: 
\begin{align}
	a_m\qty(k_1+1,k_2)\exp(-\ii2\pi\frac{l_v}{l_c}q_2+\ii\pi\sigma\frac{m}{\nu^*}+\ii\pi\frac{l_v^2}{l_c^2}k_2)&=
	a_m\qty(k_1,k_2),\label{eq:an_b1_e_}\\
	a_{m-1}\qty(k_1,k_2+1)\exp(-\ii\pi\frac{l_v^2}{l_c^2}k_1)&=a_m\qty(k_1,k_2),\ m=1,2\dots 2\nu^*-1,\label{eq:an_b2_e_}\\
	a_{2\nu^*-1}\qty(k_1,k_2+1)\exp(\ii4\pi\nu^*\frac{l_v}{l_c}q_1-\ii\pi\frac{l_v^2}{l_c^2}k_1)&=a_0\qty(k_1,k_2),\label{eq:am_0_2nu-1}
\end{align}
Similarly, the condition $\hat t_v\qty(\vbeta_i)\varphi_{\vq}\qty(\vk_c,\vk_v)=\varphi_{\vq}\qty(\vk_c,\vk_v)$ yields
\begin{align}
	a_{m}\qty(k_1+\frac{2\nu^*}{2\nu^*+\sigma},k_2)\exp(-\ii\pi k_2)&=a_m\qty(k_1,k_2),\label{eq:an_b1_v_}\\
	a_{m}\qty(k_1,k_2+1)\exp(\ii\pi\frac{l_v^2}{l_c^2}k_1)&=a_m\qty(k_1,k_2).\label{eq:an_b2_v_}
\end{align}

Repetitive applications of Eqs.~\eqref{eq:an_b1_e_} and \eqref{eq:an_b1_v_} give 
\begin{equation}
  a_m\qty(k_1+2\nu^*,k_2)=a_m\qty(k_1,k_2)\exp(\ii 4\nu^*\pi\frac{l_v}{l_c}q_2-\ii2\nu^*\pi\frac{l_v^2}{l_c^2}k_2),
\end{equation}
and
\begin{equation}
	a_m\qty(k_1+2\nu^*,k_2)=a_m\qty(k_1,k_2)\exp(\ii 2\nu^*\pi\frac{l_v^2}{l_c^2}k_2),
\end{equation}
respectively.
To be consistent, it requires
\begin{equation}
1=\exp[\ii2\pi(2\nu^{*}+1)\qty(\frac{l_{c}}{l_{v}}q_{2}-k_{2})].\label{eq:q_k_mod2}
\end{equation}

On the other hand, applying Eq.~\eqref{eq:an_b2_v_} to Eqs.~\eqref{eq:am_0_2nu-1} and \eqref{eq:an_b2_e_} gives
\begin{equation}
	a_{m+1}\qty(k_1,k_2)=a_m\qty(k_1,k_2)\exp(-\ii2\pi\frac{l_v^2}{l_c^2}k_1),\label{eq:am_recursion}
\end{equation}
and 
\begin{equation}
	a_{2\nu^*-1}\qty(k_1,k_2)=a_0\qty(k_1,k_2)\exp(\ii2\pi\frac{l_v^2}{l_c^2}k_1-\ii4\pi\nu^*\frac{l_v}{l_c}q_1).
\end{equation}
To make them consistent, it requires
\begin{equation}
1=\exp[\ii2\pi(2\nu^{\ast}+1)\qty(\frac{l_{c}}{l_{v}}q_{1}-k_{1})].\label{eq:q_k_mod1}
\end{equation}

The two consistency conditions, Eqs~\eqref{eq:am_recursion} and \eqref{eq:q_k_mod1}, impose a constraint on wavevectors:
\begin{equation}
\vk-\frac{l_{c}}{l_{v}}\vq=\frac{m_{1}\vecb_{1}+m_{2}\vecb_{2}}{2\nu^{*}+\sigma},\,m_{1},m_{2}\in\mathbb{Z}.\label{eq:k_q}
\end{equation}
Combining the constraint and the recursive relation Eq.~\eqref{eq:am_recursion}, we can express $a_m\qty(\vk)$ as: 
\begin{equation}
a_{m}\qty(\vk)=\exp(-\ii2\pi m\frac{l_{v}^{2}}{l_{c}^{2}}k_{1})\sum_{m_{1},m_{2}}c\qty(m_{1},m_{2})\delta\qty(\vq-\frac{l_{v}}{l_{c}}\vk+\frac{l_{v}}{l_{c}}\frac{m_{1}\vecb_{1}+m_{2}\vecb_{2}}{2\nu^{*}+\sigma}),\label{eq:an_delta}
\end{equation}
with $\vk$-independent coefficients $c\qty(m_1,m_2)$. 

We can obtain a recursive relation for $c\qty(m_1,m_2)$ by combining Eqs.~\eqref{eq:an_b1_e_} and \eqref{eq:an_b1_v_}:
\begin{equation}
	a_m\qty(k_1+\frac{1}{2\nu^*+\sigma},k_2)=
	a_m\qty(k_1,k_2)\exp(\ii\sigma2\pi\frac{l_v}{l_c}q_2-\ii\pi\frac{m}{\nu^*}-\ii\pi\sigma\frac{l_v^2+l_c^2}{l_c^2}k_2).
\end{equation}
It gives
\begin{equation}
	c\qty(m_1+1,m_2)=c\qty(m_1,m_2)\exp(-\ii\sigma\pi m_2\frac{l_c^2+l_v^2}{l^2}+\ii\pi\frac{l_v}{l_c}\frac{q_2}{2\nu^*+\sigma}).
\end{equation}
The coefficient can thus be written as
\begin{equation}
	c\qty(m_1,m_2)=d\qty(m_2)\exp(-\ii\pi\sigma m_1m_2\frac{l_c^2+l_v^2}{l^2}+\ii\pi\frac{l_v}{l_c}\frac{m_1q_2}{2\nu^*+\sigma}).
\end{equation}

Equation~\eqref{eq:an_b2_v_} imposes a periodicity for $d\qty(m_2)$:
\begin{equation}
	d\qty(m_2+2\nu^*+\sigma)=d\qty(m_2)\exp(-\ii\pi\frac{l_v}{l_c}q_1).
\end{equation}
We can thus construct $2\nu^*+\sigma$ independent solutions, with:
\begin{equation}
	d^{\qty(r)}\qty(m_2)=\exp(-\ii\pi\frac{l_v}{l_c}\frac{m_2q_1}{2\nu^*+\sigma})\times
	\begin{dcases}
		1&m_2= r\,\qty(\text{mod }2\nu^*+\sigma)\\
		0&\text{otherwise}
	\end{dcases},
\end{equation}
for $r=0,\cdots,2\nu^{\ast}+\sigma-1$.
Each gives an independent CF wavefunction $\varphi_{\vq}^{\qty(r)}\qty(\vk_e,\vk_v)$ for $\bm q$ defined in the reduced magnetic RPC.

Independent CF wavefunctions can also be indexed using solely a wavevector, as shown in Eq.~(10) of the main text.
To achieve this, we define
\begin{equation}
\bm{q}^{\prime}=\vq+\frac{r}{2\nu^{\ast}+\sigma}\vchi_{2}.
\end{equation}
The pair $(\bm q, r)$ with $\bm q$ defined in the reduced RPC is mapped to a point $\bm q^{\prime}$ defined in the CF-RPC (see Fig.~3 of the main text).
By making the substitution in $\varphi_{\vq}^{\qty(r)}\qty(\vk_e,\vk_v)$ and employing the identity
\begin{equation}
\phi_{\bm{q}}(\bm{x})\equiv\phi_{\bm{q}^{\prime}-\frac{r}{2\nu^{\ast}+\sigma}\bm{\chi}_{2}}(\bm{x})\propto\hat{t}\left(r\bm{\zeta}_{1}\right)\phi_{\bm{q}^{\prime}}(\bm{x}),
\end{equation}
and after some algebra, we arrive at the wavefunction Eq.~(10) of the main text, where $\bm q^{\prime}$ is re-labeled as $\bm q$.

\bibliography{ref}